\documentclass[%reprint,
groupedaddress,
preprint,
showpacs,preprintnumbers,
amsmath,amssymb,
aps,
prl
]{revtex4-1}
\usepackage{graphicx}% Include figure files
\usepackage{dcolumn}% Align table columns on decimal point
\usepackage{bm}% bold math
\usepackage{mathrsfs}
\begin{document}
\title{Anomalous superconducting proximity effect and coherent charge transport in semiconducting thin film with spin-orbit interaction}
\author{A.\,V.~Burmistrova,$^{1,2,3,4}$ I.\,A.~Devyatov,$^{2,3,4,}$}\email[]{igor-devyatov@yandex.ru}\author { I.\,E.~Batov$^{3,5}$}
\affiliation{$^{1}$Lomonosov Moscow State University, Faculty of Physics, 1(2), Leninskie gory, GSP-1, 119991 Moscow, Russia\\$^{2}$Lomonosov Moscow State University Skobeltsyn Institute of Nuclear Physics, 1(2), Leninskie gory, GSP-1, 119991 Moscow, Russia\\$^{3}$Moscow Institute of Physics and Technology, 141700 Dolgoprudny, Moscow district, Russia\\$^{4}$Moscow State Pedagogical University, 119992 Moscow, Russia\\$^{5}$Institute of Solid State Physics RAS, 142432 Chernogolovka, Moscow district, Russia}	

\date{\today}

\begin{abstract}
We present a microscopic theory of the superconducting proximity effect in a semiconducting thin film with spin-orbit interaction ($N_{SO}$) in an external magnetic field.
We demonstrate that an effective 1D Hamiltonian which describes induced superconductivity in $N_{SO}$ in contact with a usual $s$-wave superconductor possesses not only spin-singlet induced superconducting order parameter term, as commonly adopted, but spin triplet order parameter term also.
Using this new effective Hamiltonian we confirm previous results for a normal current across contacts of $N_{SO}$ with a normal metal  and for a Josephson current with the same $N_{SO}$ with induced superconductivity, obtained previously in the framework of the phenomenological Hamiltonian without spin-triplet terms.
However, a calculated current-phase relation across the transparent contact between $N_{SO}$ with induced superconductivity  in magnetic field and usual $s$-wave superconductor differs significantly from previous results.
We suggest the experiment which can confirm our theoretical predictions.
\end{abstract}


\maketitle

\begin{figure}[h]
\begin{center}
\includegraphics[width=0.82\linewidth]{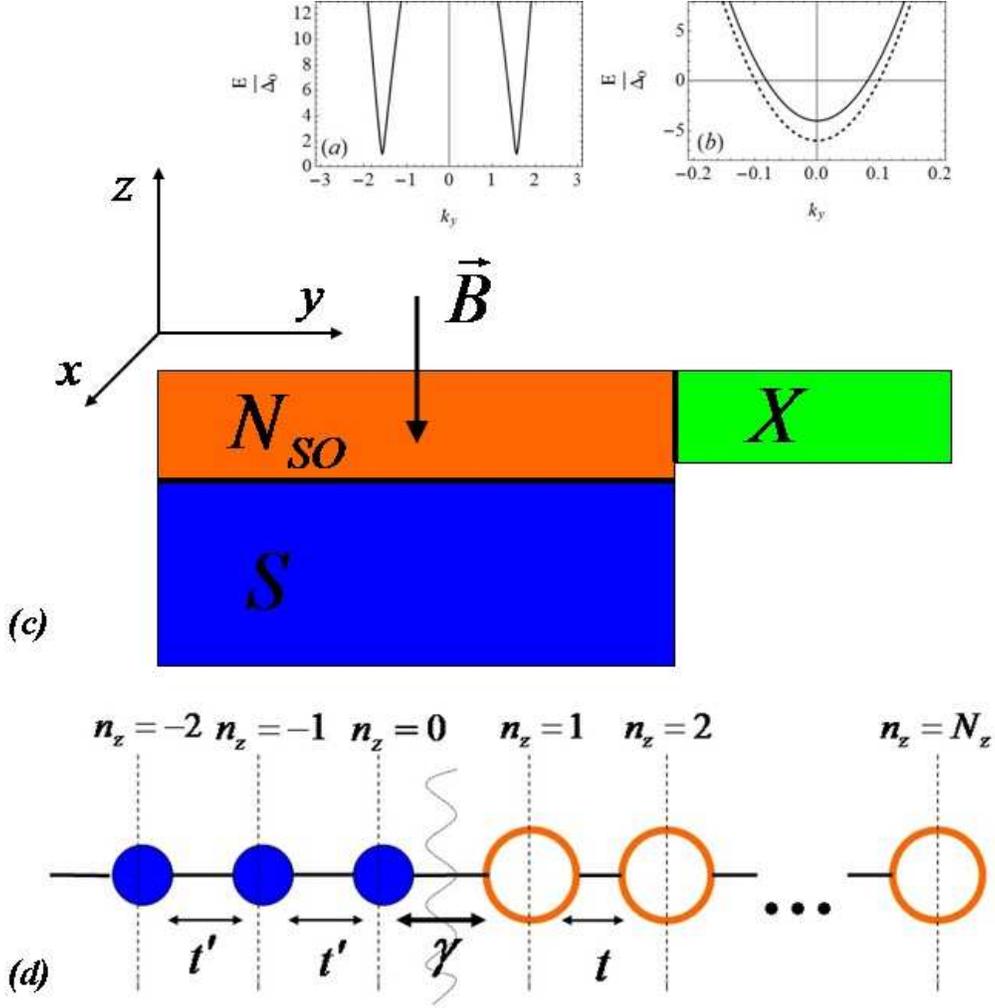}
\caption{The quasiparticle excitation spectrum for $S$ (a); for $N_{SO}$ in $z$ direction  (b); schematic illustration of the model under consideration of the
$S/N_{SO}$ and $N_{SO}/X$ junctions, where $X$ is $N,S$ or $N_{SO}$ with induced superconductivity  (c); tight-binding model of the proximity effect in $z$-direction for $S/N_{SO}$ junction (d).}
\label{Ris1}
\end{center}
\end{figure}

How is it possible to describe the superconductivity induced in materials without attractive electron-electron interaction?
The answer is well-known for the case of a contact between a normal metal and a usual $s$-wave spin-singlet superconductor ($S$) \cite{vol,mcm}.
In this case the induced superconductivity in the metal without electron-electron attraction can be described by almost the same Hamiltonian as in the usual superconductor \cite{vol}.
However, the type of the effective Hamiltonian which can describe the induced superconductivity in unusual materials,
such as semiconducting thin film ($N_{SO}$) and topological insulators, is less evident.
These structures - usual $s$-wave superconductor in contact with the semiconducting thin film ($S/N_{SO}$ contact) or topological insulator (Fig.\ref{Ris1}(c)) - are very interesting from both fundamental and practical points of view due to the possibility to find in them zero-energy Majorana mode.
These zero-energy modes can be interpreted as Majorana quasiparticles, which are their own antiparticles, in contrast to the usual Fermi particles when particles and antiparticles are distinct \cite{alic}.
Majorana quasiparticles obey non-Abelian statistics, rather than Fermi and Dirac statistics, which is very unorthodox for solid state physics \cite{nay}.
Majorana quasiparticles are promising candidates for using in fault tolerant quantum computations \cite{kit}.
The knowledge of the correct effective Hamiltonian taking into account the induced superconductivity in $N_{SO}$ permits to plan the crucial experiment to detect Majorana quasiparticles
as well as to provide the future investigations of the 	element base of the  fault tolerant quantum computer.

Surprisingly, despite the great interest to the investigation of the perspective heterostructures which can contain Majorana quasiparticles, there is still no consistent derivation of the effective Hamiltonian taking into account the influence of the superconductor and finite geometrical size of unconventional materials, as it was done for the contact of the normal metal with the usual superconductor \cite{vol}. Even microscopical self-consistent calculations of the proximity effect in such structures \cite{black1, black2,lind1,lind2,maslov} were resricted to find just averaged pairing amplitudes, i.e. Green function components, but not an effective Hamiltonian.
Existing attempts of derivation of the effective Hamiltonian \cite{alic, tewari,pot} describe the connection between the usual $s$-wave superconductor and the topological material in terms of the tunnel  Hamiltonian in the momentum space which does not permit to take into account the finite width of the topological material layer, coherent reflections between boundaries as well as scattering between spin bands.
Naturally, the important features of the effective Hamiltonian as triplet pairing component and momentum dependence of pairing component were missed.

In this Letter, based on our tight-binding approach \cite{new,epl1,prb1}
we investigate the superconducting  proximity effect in the metal with spin-orbit interaction from the usual $s$-wave superconductor.
We demonstrate that the effective 1D Hamiltonian in the basis $\Psi=(\Psi^{SO,\uparrow},\Psi^{SO,\downarrow},\bar{\Psi}^{SO,\uparrow},\bar{\Psi}^{SO,\downarrow})$ should have the following form:

\begin{equation}
\widehat{H}_{eff}=\begin{pmatrix}
\xi-h & \lambda k_y & \Delta_1(k_y) & \Delta_2(k_y) \\
\lambda k_y & \xi+h & -\Delta_2(k_y) & \Delta_3(k_y) \\
\Delta_1(k_y) & -\Delta_2(k_y) & -\xi+h & \lambda k_y \\
\Delta_2(k_y) & \Delta_3(k_y) & \lambda k_y & -\xi-h
\end{pmatrix}
\label{ham}
\end{equation}

\noindent instead of widely used Hamiltonian \cite{alic, tewari,pot} without triplet terms $\Delta_1(k_y)=\Delta_3(k_y)=0$ and without momentum dependence of the induced singlet order parameter $\Delta_2(k_y)=const$. In Eq.(\ref{ham}) $\xi$ is a single particle excitation energy, 
$h$ is the Zeeman energy related to the magnetic field $B$ applied in the $z$ direction  
(Fig.\ref{Ris1}(c)), $h=g_e^* \mu_B B/2$, $g_e^*$ is the Land\'{e} factor, $\mu_B$ is the Bohr magneton, $\lambda$ is a spin-orbit constant (we consider the Rashba model \cite{rash}) and $k_y$ is a momentum parallel to the interface.
Using this effective Hamiltonian (\ref{ham}) we demonstrate that it leads to well-known results for normal current for the contact with a normal metal ($N/S_{SO}$  junction)  \cite{stan1,flen,law}  and for the Josephson current for the symmetric junction with the same metal with induced superconductivity ($S_{SO}/c/S_{SO}$ Josephson junction) \cite{lut}.
However, we demonstrate that the use of the Hamiltonian (\ref{ham}) leads to very unusual current-phase relations for the contact of this heterostructure (Fig.\ref{Ris1}(c)) with usual $s$-wave superconductor ($S/c/S_{SO}$ Josephson junction).

We consider the $S/N_{SO}$ heterostructure which is depicted in Fig.\ref{Ris1}(c,d).
We suppose that  $S/N_{SO}$ boundary is sufficiently smooth, so the  momentum parallel to the interface is conserved.
The wave function in $S$ material corresponding to the case of the bound states with $E<\Delta_0$ has the usual form \cite{btk}.
In $N_{SO}$ spin-orbit interaction couples spins with different directions with each other, which leads to the wave functions in the form of the superposition of eight bispinors \cite{Tanaka2006}.

To solve the problem of the induced superconductivity in $N_{SO}$ we should match wave functions on boundaries. In the tight-binding approximation it is suitable to use the  boundary conditions for $S/N_{SO}$ interface at $n_z=0$ \cite{new}. $n_z$ means the number of atoms in the tight-binding scheme (Fig.\ref{Ris1}(d)),
which corresponds to the coordinate $z$ by the following relation $z=a \cdot n_z$, where $a$ is the distance between atoms.
For simplicity we put $a=1$ in the remaining part of this Letter.
Open boundary of $N_{SO}$ to vacuum corresponds to $n_z=N_z$.
For spin-up components of the wave function these boundary conditions have the following form:

\begin{equation}
\left\{
\begin{aligned}
&t'\Psi^{S,\uparrow}_1=\gamma \Psi^{SO,\uparrow}_1 ,\\
&t'\bar{\Psi}^{S,\uparrow}_1=\gamma \bar{\Psi}^{SO,\uparrow}_1, \\
&t\Psi^{SO,\uparrow}_0=\gamma \Psi^{S,\uparrow}_0 ,\\
&t\bar{\Psi}^{SO,\uparrow}_0=\gamma \bar{\Psi}^{S,\uparrow}_0 ,
\end{aligned}
\right.\label{bc}
\end{equation}

\noindent where $t$, $t'$, $\gamma$ are tight-binding hopping amplitudes in $N_{SO}$, $S$ and across boundary, respectively (Fig.\ref{Ris1}(d)). For spin-down components boundary conditions are similar to Eq. (\ref{bc}),  and
for open boundary at $n_z=N_z$ one has the following boundary conditions:
$\Psi^{SO,\uparrow}_{N_z}=
\Psi^{SO,\downarrow}_{N_z}=
\bar{\Psi}^{SO,\uparrow}_{N_z}=
\bar{\Psi}^{SO,\downarrow}_{N_z}=0$. In these boundary conditions, $\Psi^{X,\uparrow(\downarrow)}_{n_z}$ corresponds to the electron component of the wave function in $X$ with spin up(down) on the atom with number $n_z$, $\bar{\Psi}^{X,\uparrow(\downarrow)}_{n_z}$ corresponds to the hole component of the wave function in $X$ with spin up(down) on the atom with number $n_z$, where $X$ is $N_{SO}$ or $S$.

Substitution of the wave functions to the boundary conditions Eq.(\ref{bc}) leads to the transcendental equation, which solution allows to obtain the induced excitation spectrum in $N_{SO}$. The obtained induced excitation spectra are rather similar to the previously obtained results with the phenomenological Hamiltonian \cite{sau2,snel1}. For the case without magnetic field and for values of the magnetic field smaller than critical there are two gaps in the excitation spectrum: the first gap corresponds to the smaller value of $k_y$ and the second gap corresponds to the larger value of $k_y$. At critical value of the Zeeman field $h=h_c$ the first gap is closed, and then for values of the magnetic field larger than critical the first gap is reopened.

However, the Majorana states can arise at the end of the clean $N_{SO}$ \cite{lut,oreg}. Therefore, the investigation of the transport in $y$-direction of $N_{SO}$ (Fig.\ref{Ris1}(c)) is of great interest. The most common way to do it is to construct the effective 1D Hamiltonian using obtained wave functions. For this purpose one needs to construct the Green function for the lowest subband of $N_{SO}$ \cite{vol} and then find the effective 1D Hamiltonian from the equation $-\hat{H}_{eff}(k_y)\hat{G}(k_y)=1$. The components of the retarded Green function presented by $4\times4$ matrix are expressed through the components of the wave functions $\Psi^{(\alpha)}_{n_z}(E_i(k_y),k_y)$ \cite{vol}:

\begin{equation}
\begin{aligned}
&&G^{R,(\alpha\beta)}_{n_z,n_z}(k_y)=\sum_{i=1,...,4}\frac{\langle\Psi^{(\alpha)}_{n_z}(E_i(k_y),k_y)\Psi^{(\beta)*}_{n_z}(E_i(k_y),k_y)\rangle}{E+i0-E_{i}}
\end{aligned}
\label{fGreen2}
\end{equation}

\noindent where $\alpha,\beta=1,2,3,4$, the sum is taken over all four branches of the induced spectrum, the brackets denote averaging over $n_z  (1\leqslant n_z\leqslant N_z)$. The obtained effective 1D Hamiltonian is presented by Eq.(\ref{ham}) with nonzero triplet terms $\Delta_1(k_y)$ and $\Delta_3(k_y)$.

\begin{figure}
\begin{minipage}[h]{0.47\linewidth}
\center{\includegraphics[width=1\linewidth]{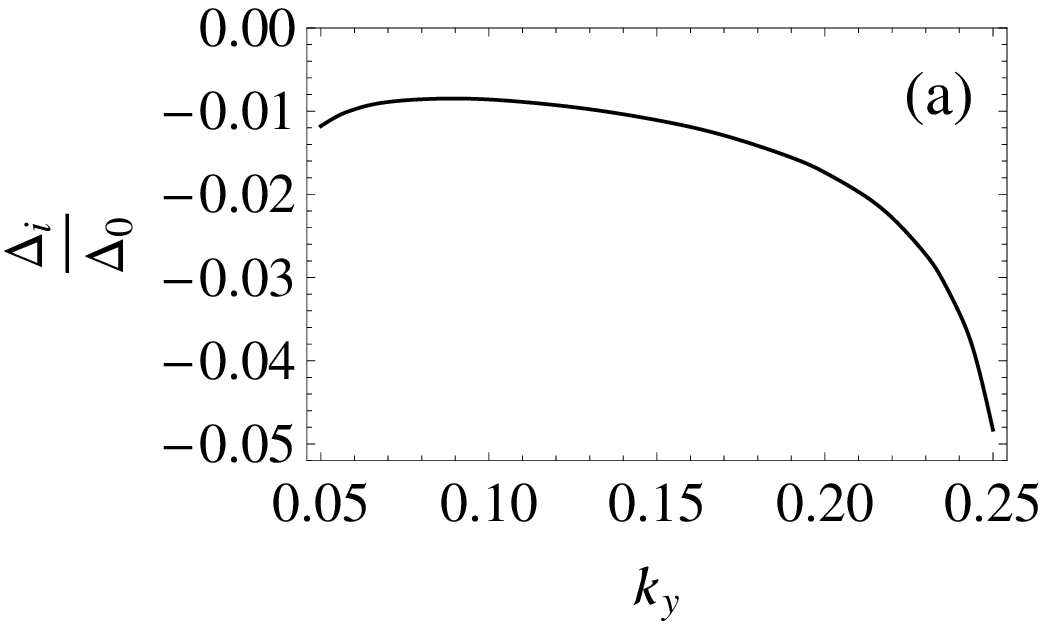}} \\
\end{minipage}
\hfill
\begin{minipage}[h]{0.47\linewidth}
\center{\includegraphics[width=1\linewidth]{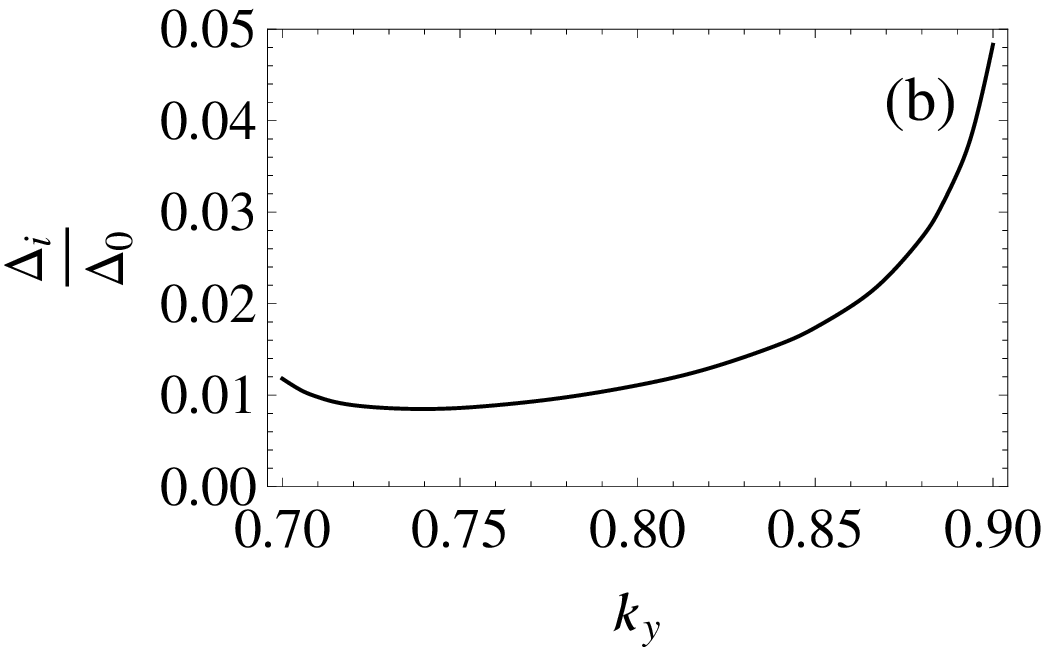}} \\
\end{minipage}
\caption{The dependencies of the induced triplet and singlet superconducting order parameters (a) $\Delta_1(k_y)=-\Delta_3(k_y)=-\Delta_2(k_y)$ near the first gap and (b) $\Delta_1(k_y)=-\Delta_3(k_y)=\Delta_2(k_y)$ near the second gap on the wave vector for zero value of the magnetic field $B=0$.}
\label{Delta_h0}
\end{figure}

\begin{figure}
\begin{minipage}[h]{0.47\linewidth}
\center{\includegraphics[width=1\linewidth]{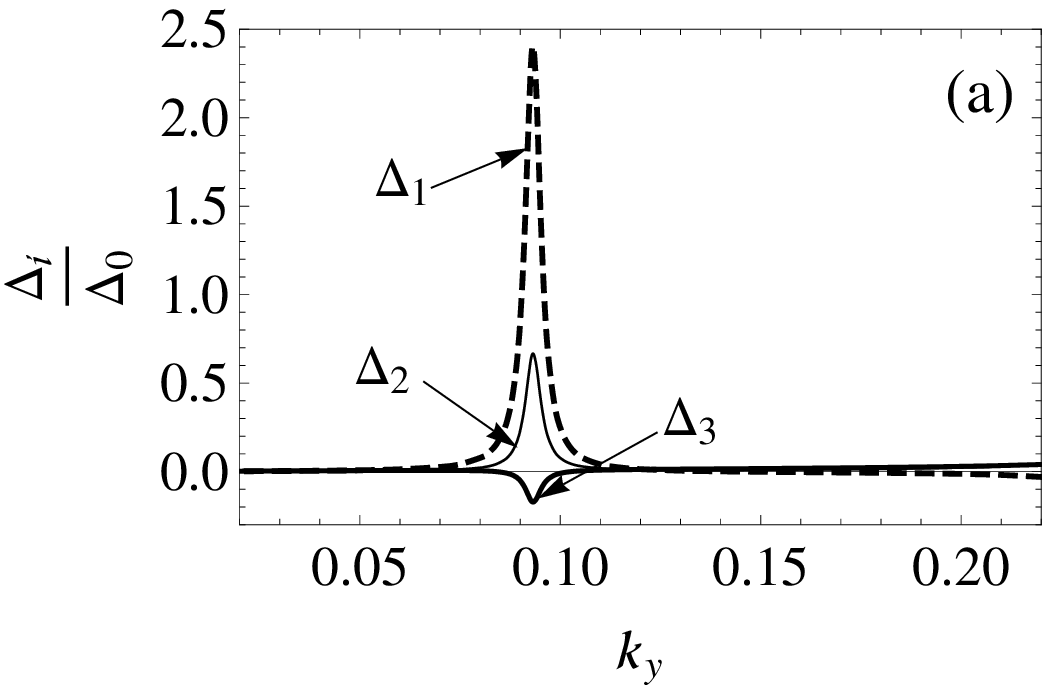}} \\
\end{minipage}
\hfill
\begin{minipage}[h]{0.49\linewidth}
\center{\includegraphics[width=1\linewidth]{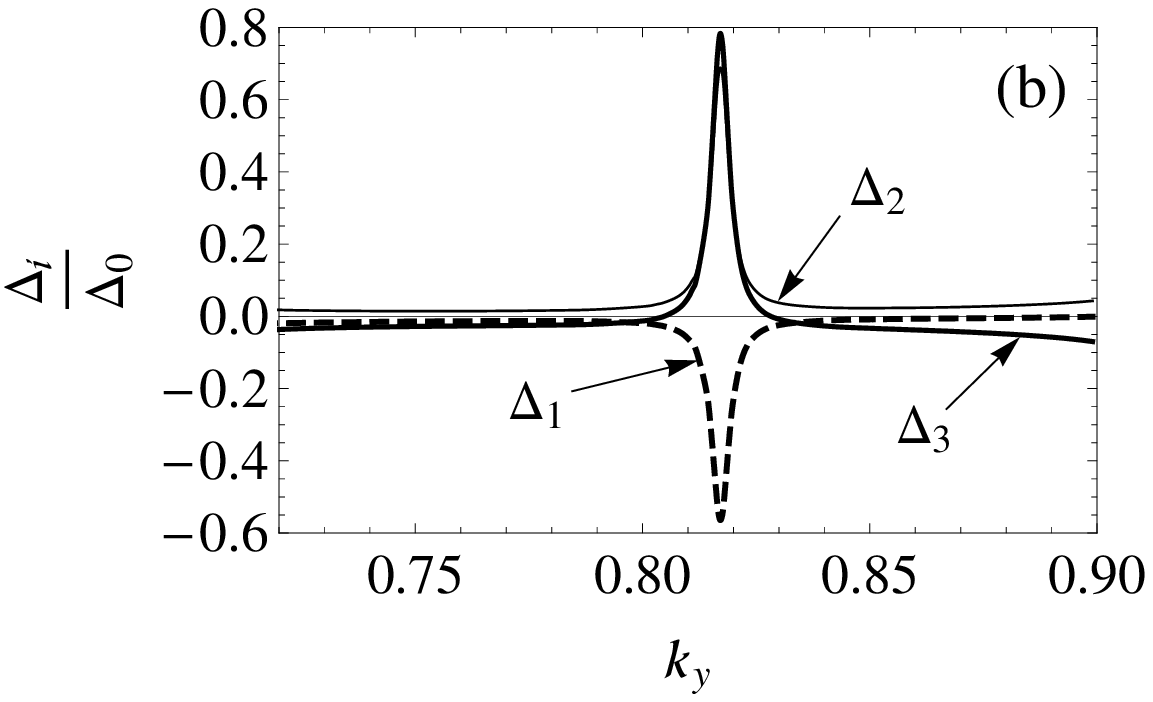}} \\
\end{minipage}
\caption{The dependencies of the induced triplet and singlet superconducting order parameters (a) near the first gap and (b)  near the second gap on the wave vector for the value of the Zeeman field $h=\Delta_0$ smaller than critical value. Solid thick line corresponds to the dependence on $k_y$ of the triplet component $\Delta_3$, dashed line - the triplet component $\Delta_1$ and solid thin line - the singlet component $\Delta_2$.}
\label{Delta_h1}
\end{figure}

\begin{figure}
\begin{minipage}[h]{0.48\linewidth}
\center{\includegraphics[width=1\linewidth]{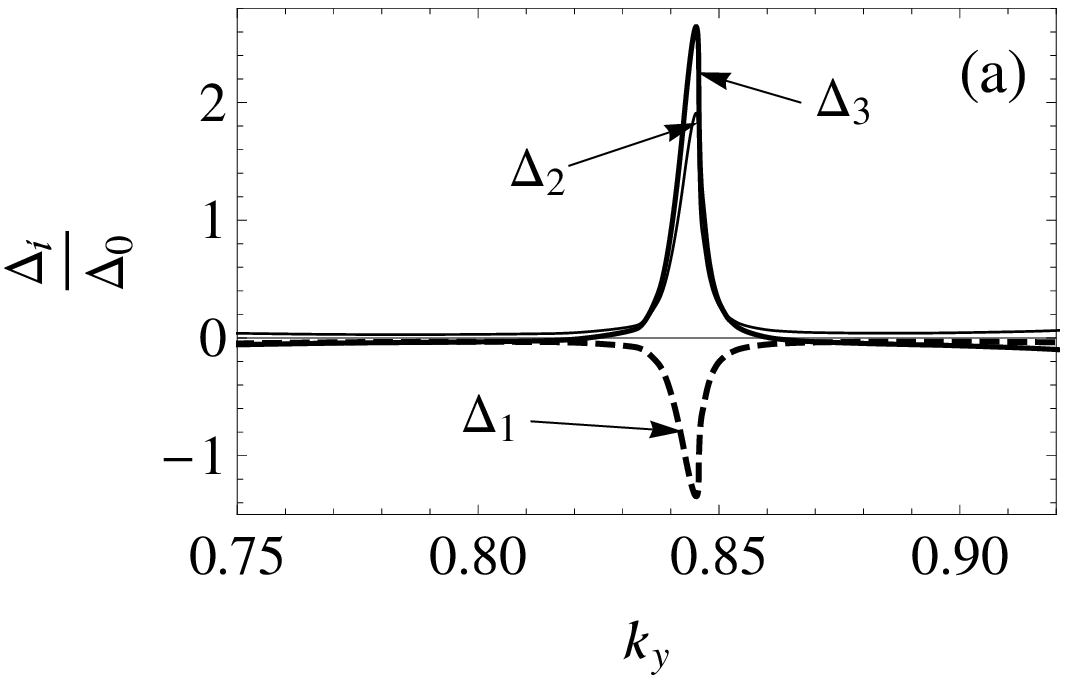}} \\
\end{minipage}
\hfill
\begin{minipage}[h]{0.47\linewidth}
\center{\includegraphics[width=1\linewidth]{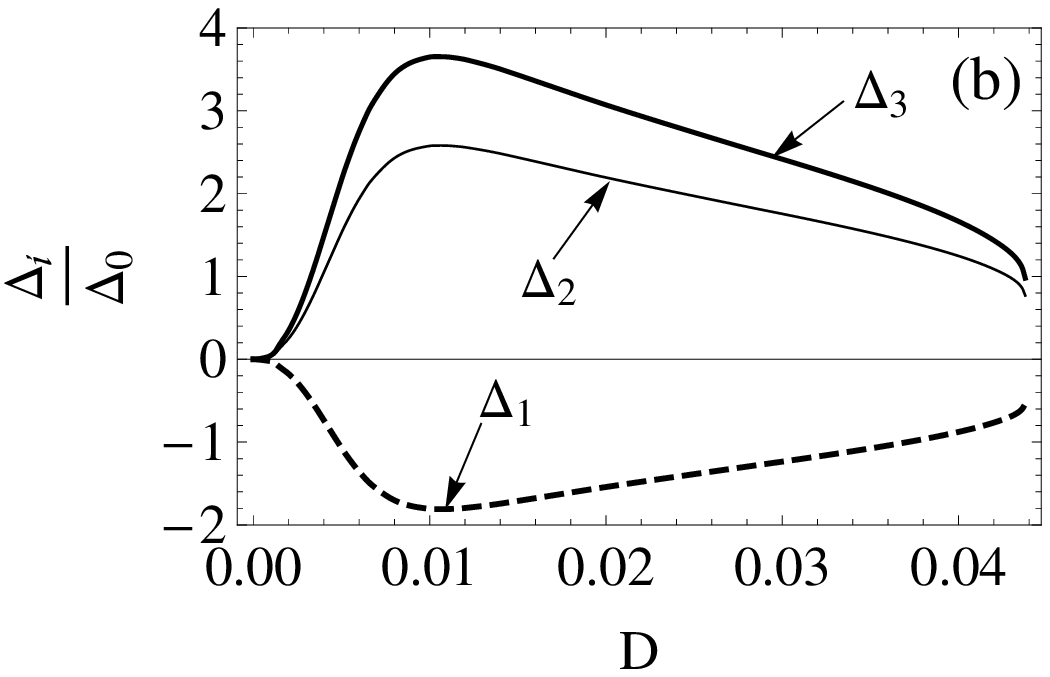}} \\
\end{minipage}
\caption{(a) The dependencies of the induced triplet and singlet superconducting order parameters near the second gap on the wave vector for the value of the Zeeman field $h=2\Delta_0$ larger than critical value. (b) The dependencies of the induced triplet and singlet superconducting order parameters near the second gap on the transparency of the interface $D$ for the value of the Zeeman field $h=2\Delta_0$ larger than critical value for $k_y=0.845$, which corresponds to the second gap. Solid thick line corresponds to the  triplet component $\Delta_3$, dashed line - the triplet component $\Delta_1$ and solid thin line - the singlet component $\Delta_2$. }
\label{Delta_h2}
\end{figure}

\begin{figure}
\begin{minipage}[h]{0.47\linewidth}
\center{\includegraphics[width=1\linewidth]{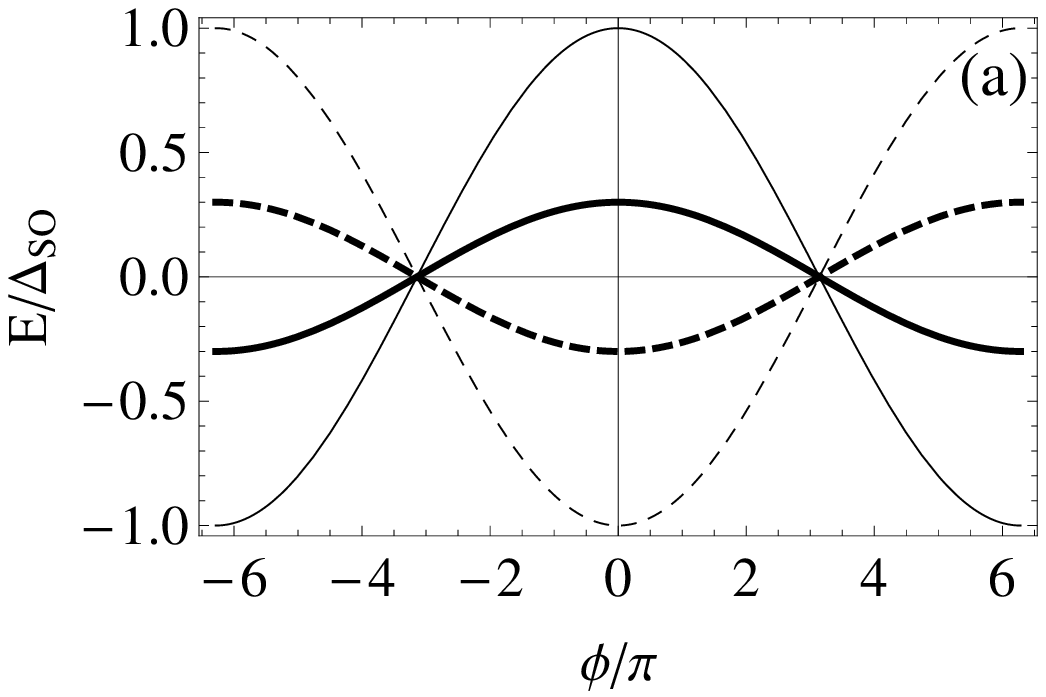}} \\
\end{minipage}
\hfill
\begin{minipage}[h]{0.47\linewidth}
\center{\includegraphics[width=1\linewidth]{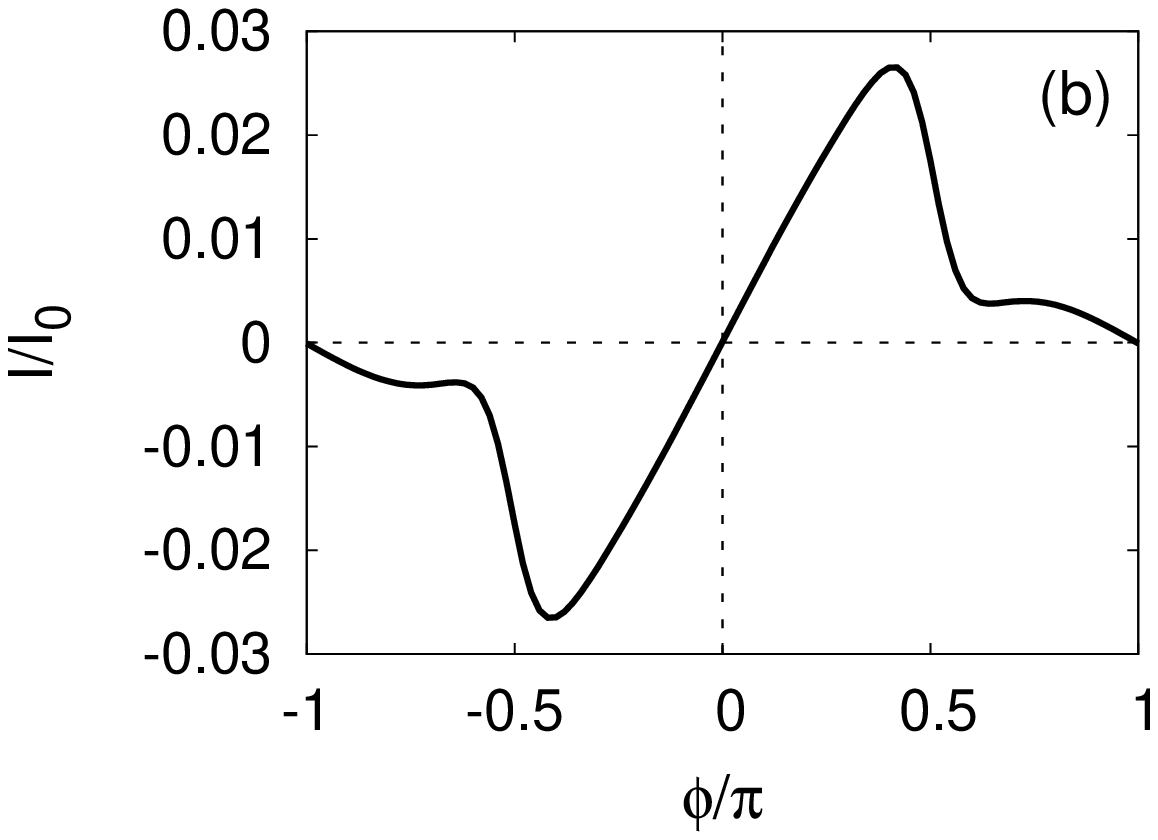}} \\
\end{minipage}
\vfill
\begin{minipage}[h]{0.47\linewidth}
\center{\includegraphics[width=1\linewidth]{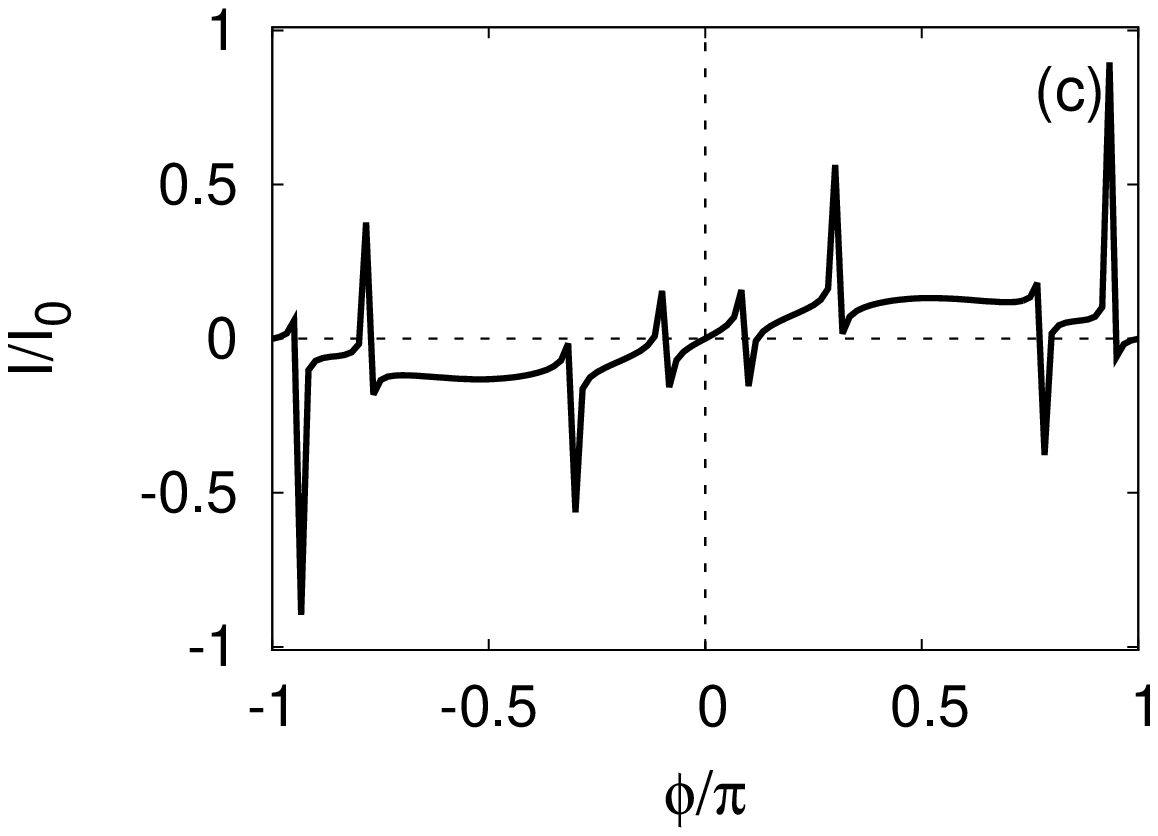}} \\
\end{minipage}
\hfill
\begin{minipage}[h]{0.47\linewidth}
\center{\includegraphics[width=1\linewidth]{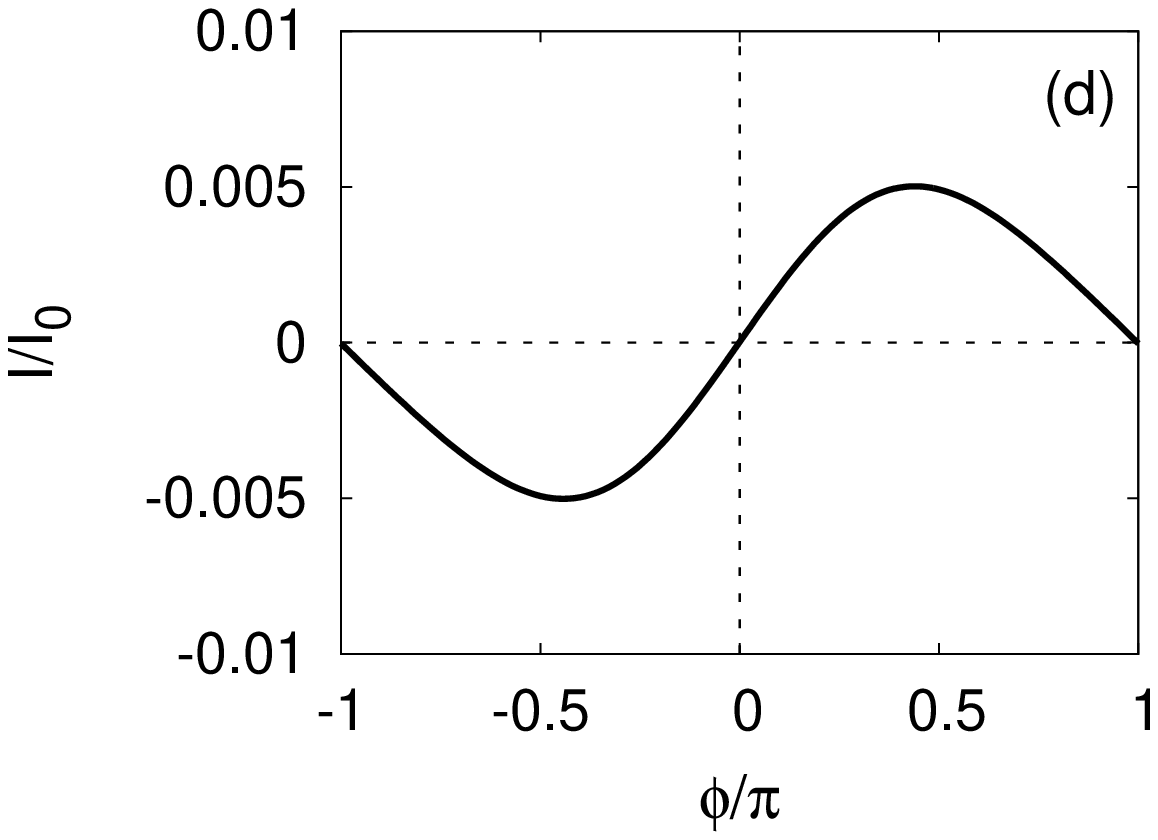}} \\
\end{minipage}
\caption{ (a) Andreev bound states for $S_{SO}/c/S_{SO}$ junction; thick solid and dashed lines correspond to the transparency of the interface $D=0.145$, thin solid and dashed lines - to the $D=1$; (b) the current-phase relation for the $S/c/S_{S0}$ junction for the following values of parameters $t=t_S=-200\Delta_0, \lambda=18\Delta_0,\mu=398\Delta_0,h=5\Delta_0,k_S=2.5,\gamma=-200\Delta_0,D=0.135,I_0=2e\Delta_{SO}/\hbar$; (c) the current-phase relation for the $S/c/S_{S0}$ junction for the same values of parameters as in (b) but $k_S=0.25,\gamma=-180\Delta_0,D=0.83$; (d)  the current-phase relation for the $S/c/S_{S0}$ junction for the same values of parameters as in (b) but $k_S=0.25,\gamma=-80\Delta_0,D=0.05$.}
\label{Ifi}
\end{figure}

The dependencies of the induced triplet and singlet superconducting order parameters on the wave vector for zero magnetic field are depicted in Fig.\ref{Delta_h0}.
We assume that the chemical potential $\mu$ in this and the following figures  has the following value: $\mu=1215\Delta_0$.
Even for this case $B=0$ the induced triplet components are nonzero, have significant $k_y$ dependencies and satisfy the relation $\Delta_1(k_y)=-\Delta_3(k_y)=-\Delta_2(k_y)$ near the first gap and $\Delta_1(k_y)=-\Delta_3(k_y)=\Delta_2(k_y)$ near the second gap. The nonzero triplet components arise due to the presense of the spin-orbit interaction in the layer of $N_{SO}$ with finite thickness where coherent reflections on the boundaries exist.

The dependencies of the induced triplet and singlet superconducting order parameters on the wave vector  for the value of the Zeeman field $h=\Delta_0$
%(here  we put the chemical potential $\mu=0$, assumed that it tuned by the gate)
smaller than critical are depicted in Fig.\ref{Delta_h1}. Solid thick line corresponds to the dependence on $k_y$ of the triplet component $\Delta_3$, dashed line - the triplet component $\Delta_1$ and solid thin line - the singlet component $\Delta_2$. From Fig.\ref{Delta_h1}(a) one can see that near the first gap the magnitude of the triplet component $\Delta_1(k_y)$ is significantly larger than the magnitude of the triplet component $\Delta_3(k_y)$, but near the second gap, as one can see from Fig.\ref{Delta_h1}(b), their magnitudes are approximately equal.

The dependencies of the induced triplet and singlet superconducting order parameters near the second gap on the wave vector for the value of the Zeeman field $h=2\Delta_0$ larger than critical value are depicted in Fig.\ref{Delta_h2}(a), and the dependencies of the induced triplet and singlet superconducting order parameters near the second gap on the transparency of the interface are depicted in Fig.\ref{Delta_h2}(b) for the same value of the magnetic field for $k_y=0.845$ which corresponds to the second gap. From this figure one can see that for almost all values of transparency of the interface the magnitudes of the components of the induced order parameter are larger than the magnitude of the order parameter $\Delta_0$ in $S$. However, it is possible to demonstrate that the gap in the induced spectrum is always smaller than $\Delta_0$.

Using the  effective 1D Hamiltonian (\ref{ham}) we have calculated the current-voltage characteristics (IVC) of $N/S_{SO}$ junction in $y$ direction (Fig.\ref{Ris1}(c)) in the framework of the approach \cite{new}. We demonstrate that at high Zeeman field $h>h_c$ zero-energy singularity in the IVC appears, which can be interpreted as zero-energy Majorana states.
Thus, our results do not contradict to the previous results \cite{stan1,flen,law}, where Majorana states and corresponding zero-energy singularities in the IVC of $N/S_{SO}$  junction were predicted.

However, the crucial experiment to determine the surface bound states is the Josephson tunneling experiment.
With the aim to plan such experiment we have calculated Andreev bound states and the Josephson current in different short (the length of the junction is much smaller than the coherence length) superconducting junctions containing $S_{SO}$.

We have calculated  Andreev bound states for symmetric Josephson $S_{SO}/c/S_{SO}$ junction, which are presented in
Fig.\ref{Ifi}(a).
Thick solid and dashed lines in
Fig.\ref{Ifi}(a) correspond to the transparency of the interface $D=0.145$, thin solid and dashed lines  correspond to the transparency of the interface $D=1$.
It follows from
Fig.\ref{Ifi}(a)
that $4 \pi$ periodicity of  Josephson current-phase relation exists for any values of the transparency of the interface.
Thus, our calculations confirm previous results for the Josephson current in symmetric superconducting $S_{SO}/c/S_{SO}$ junctions
\cite{lut} which were obtained using a phenomenological Hamiltonian without triplet terms $\Delta_1(k_y)=\Delta_3(k_y)=0$ and without momentum dependence of the induced singlet order parameter $\Delta_2(k_y)=const$ in Eq. (\ref{ham}).

However, an investigation of the current-phase relation of \emph{asymmetric} short Josephson junctions, one bank of which is $S_{SO}$, and another bank is $S$, provides the possibility to  distinguish between a phenomenological and microscopically obtained Hamiltonians.
In Fig.\ref{Ifi}(b)-(d) current-phase relations of asymmetric $S/c/S_{S0}$ Josephson junctions calculated from Hamiltonian (\ref{ham}) are presented for different values of $S/c/S_{S0}$ interface transparency.
One can see that for relatively large values of $S/S_{S0}$ interface transparency ($D=0.135$ at Fig.\ref{Ifi}(b) and $D=0.83$ at Fig.\ref{Ifi}(c)) calculated current-phase relations are rather unusual and significantly differ from well-known dependencies.
Current-phase relation of $S/c/S_{S0}$ Josephson  junction with small $S/S_{S0}$ interface transparency (Fig.\ref{Ifi}(d)) demonstrates usual sinusoidal dependence.
Therefore, an investigation of current-phase relations of quite transparent asymmetric $S/c/S_{S0}$ Josephson  junctions
provides a possibility to verify our results.

\begin{figure}[h]
\begin{center}
\includegraphics[width=0.75\linewidth]{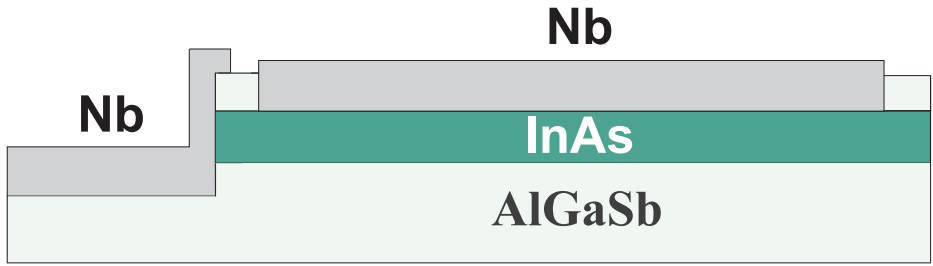}
\caption{A possible experimental realization of a ballistic asymmetric $S/c/S_{S0}$ Josephson junction}
\label{Scheme}
\end{center}
\end{figure}

We now discuss the design of an asymmetric $S/c/S_{S0}$ Josephson junction and a possible experimental setup for measuring the predicted effects. The Josephson junction we propose is based on a high-mobility InAs/AlGaSb heterostructure and niobium electrodes \cite{Eroms,Rohlfing}. A schematic cross section of the hybrid Nb-InAs/AlGaSb structure is shown in Fig.\ref{Scheme}. The hybrid nanostructure is defined by electron beam lithography, selective reactive ion etching, and Nb sputter deposition. Only the top AlGaSb-layer is etched in the central part of the structure, while the InAs-channel continues underneath the niobium \cite{Eroms,Rohlfing}. An etched semiconductor mesa (with an Nb layer on top) is laterally contacted to the superconducting niobium lead \cite{Amado}.
Highly transparent contacts can be formed in the junction by exploiting an Ar plasma cleaning of the contact area prior to the Nb sputter deposition \cite{Eroms,Rohlfing,Amado}.
The mean free path in the InAs-quantum well $l_e>3\mu m$ \cite{Eroms,Rohlfing}, allowing for ballistic transport in nanostructures. The current-phase relation of the assymmetric Josephson junction can be determined by incorporating the junction into a superconducting loop coupled to a dc SQUID, allowing measurement of the junction phase difference \cite{Frolov,eng}.

In summary, we present here a microscopic theory of the superconducting proximity effect in the contact of usual $s$-wave superconductor with a metal with spin-orbit interaction in an  applied magnetic field. Our theory takes into account scattering between spin bands at the boundaries and finite size of the metal, which were missed in the previous investigation of the proximity effect in such structures \cite{alic, tewari,pot}. We obtain the effective 1D Hamiltonian (\ref{ham}) which describes the induced superconductivity in such metal and demonstrates the presence of the spin-triplet order parameter components in it which contradicts to previous investigations where only spin-singlet component were obtained  \cite{alic, tewari,pot}.
Nevertheless, using the effective Hamiltonian (\ref{ham}) does not frustrate the main results obtained previously for such materials: we confirm the existence of zero-energy bound state on the boundary of this material with a normal metal \cite{stan1,flen,law} and a $4 \pi$ periodicity of the Josephson current in a symmetric junction \cite{lut}.
At the same time, we show that the Josephson current-phase dependence of quite  transparent contact of this material with conventional $s$-wave superconductors in magnetic fields is rather unusual.
We suggest an experiment which can confirm the existence of triplet pairing terms in 1D effective Hamiltonian of a $N_{SO}$ in contact with usual superconductor.

\begin{acknowledgments}
We thank T.M. Klapwijk, Y. Asano and A.A. Golubov for useful discussions.
A.V.B and I.A.D. acknowledge
financial  support from the Russian Foundation for Basic
Research, projects N 13-02-01085, 14-02-31366-mol-a, 15-52-50054 
and
financial  support from the Ministry of Education and Science of the Russian Federation, contract N 14.B25.31.0007,
I.E.B. acknowledges
financial  support from the Russian Foundation for Basic
Research, project 14-02-01041  and
financial  support from the Ministry of Education and Science of the Russian Federation, contract N 14.Y26.31.0007
\end{acknowledgments}

%\bibliography{biblio}

\end{document}